\documentclass[11pt]{article}
\usepackage[utf8]{inputenc}
\usepackage{geometry}
\usepackage{authblk}
\usepackage{amssymb}
\usepackage{amsmath}
\usepackage{amsthm}
\allowdisplaybreaks[4]
\usepackage{slashed}
\usepackage{graphicx,color,epsfig}
\usepackage{multirow}
\usepackage{cite}
\usepackage[colorlinks=true,linkcolor=red,citecolor=blue]{hyperref}
\begin{document}
\renewcommand{\arraystretch}{0.5}
\newcommand{\beq}{\begin{eqnarray}}
\newcommand{\eeq}{\end{eqnarray}}
\newcommand{\non}{\nonumber\\ }
\newcommand{\acp}{ {\cal A}_{CP} }
\newcommand{\psl}{ p \hspace{-1.8truemm}/ }
\newcommand{\nsl}{ n \hspace{-2.2truemm}/ }
\newcommand{\vsl}{ v \hspace{-2.2truemm}/ }
\newcommand{\epsl}{\epsilon \hspace{-1.8truemm}/\,  }
\newcommand{\tf}{\textbf}
\title{Study of Quasi-two-body $B_{(s)}\to \phi (f_0(980)/f_2(1270)\to)\pi\pi$ Decays in Perturbative QCD Approach}
\author[1]{Zhi-Tian Zou}
\author[1]{Lei Yang}
\author[1,2]{Ying Li$\footnote{liying@ytu.edu.cn}$}
\author [3]{Xin Liu}
\affil[1]{\it \small Department of Physics, Yantai University, Yantai 264005, China}
\affil[2]{\it \small Center for High Energy Physics, Peking University, Beijing 100871, China}
\affil[3]{\it \small Department of Physics, Jiangsu Normal University, XuZhou 221116,China}
\maketitle
\vspace{0.2cm}
\begin{abstract}
 In 2017, LHCb collaboration reported their first observation of the rare decays $B_s \to \phi (f_0(980)$ $/f_2(1270) \to ) \pi^+\pi^-$ and the evidence of $B^0 \to \phi(f_0(980)/f_2(1270)\to)\pi^+\pi^-$. Motivated by this, we study these quasi-two-body decays in the perturbative QCD approach. The branching fractions, $CP$ asymmetries and the polarization fractions are calculated. We find that within the appropriate two-meson wave functions, the calculated branching fractions are in agreement with the measurements of LHCb. Based on the narrow-width approximation, We also calculate the branching fractions of the quasi-two-body $B_{d,s}\to \phi(f_0(980)/f_2(1270)\to) \pi^0\pi^0$ and $B_{d,s}\to \phi(f_2(1270)\to)  K^+K^-$, and hope the predictions to be tested in the ongoing LHCb and Belle II experiments. Moreover, the processes $B_{d,s}\to \phi f_2(1270)$ are also analyzed under the approximation. We note that the $CP$ asymmetries of these decays are very small, because these decays are either penguin dominant or pure penguin processes.
\end{abstract}
\newpage
\section{Introduction}
As two-body decays, three-body non-leptonic $B$ mesons decays also provide us an opportunity to test the factorization hypothesis adopted in studying two-body hadronic decays, to test the Standard Model (SM) by measuring the CKM matrix parameters such as the phases $\alpha$ and $\gamma$, and to understand the mechanism of the $CP$ violation especially the sources of the strong phases. In the experimental side, many three-body $B$ meson decays have been analyzed extensively by the BaBar \cite{Aubert:2003mi, Aubert:2005ce, Aubert:2009me, Aubert:2009av, Aubert:2007bs, Aubert:2006nu, Aubert:2007sd, Lees:2011nf, Lees:2012kxa}, Belle \cite{Garmash:2006fh, Garmash:2005rv, Garmash:2004wa, Dalseno:2008wwa, Garmash:2003er, Abe:2002av, Nakahama:2010nj},CLEO \cite{Eckhart:2002qr}, and LHCb \cite{Aaij:2014iva, Aaij:2013bla, Aaij:2013sfa, Aaij:2018rol ,Aaij:2016qnm, Aaij:2019nmr, Aaij:2017zgz, Aaij:2019jaq} collaborations. Such abundant experimental analyses promote the theoretical studies. Unlike the two-body $B$ decays where the kinematics are fixed, in three-body decays the momentum of each final state is variable, and both resonant and non-resonant contributions are involved. Therefore, how to distinguish the resonant and non-resonant contributions reliably is a key part in studying the three-body decays. As an effective approach, Dalitz plot is particularly adopted to analyze the three-body decays, and the phase space can be divided into different regions with special kinematical configurations. The centre of the Dalitz plot indicates that all three final particles have a large energy ($E\sim M_B/3$) in the $B$ meson rest frame and none of them flies collinearly to any others. The edge of the Dalitz plot corresponds to the kinematical configuration where the two mesons fly collinearly, generating an invariant mass recoiling against the third bachelor meson. The three corners represent the kinematical configurations where one of the final mesons is soft or even rest, and the other two particles move back-to-back with large energy ($E \sim M_B/2$). It is accepted by most of us that the contributions at the centre of the Dalitz plot is $\alpha_s$ suppressed relative to that at edge \cite{Virto:2016fbw}. In oder to seize the main contributions, we focus on the physics at the edge of the Dalitz plot, where the two collinear mesons move almost in the same direction and can be viewed as a cluster, where two moving particles might form many resonances with different angular momenta. These decays are the so-called quasi-two-body decays beyond the narrow-width approximation \cite{Virto:2016fbw}.

So far, the factorization in nonleptonic three-body $B$ decays has not been rigorously proved, and all theoretical studies are still model dependent. In order to analyze the amplitude in experiments, many popular methods have been adopted, such as the isobar model \cite{Sternheimer:1961zz,Herndon:1973yn}, the $K$-matrix formalism \cite{Chung:1995dx}, and the quasi-model-independent analysis \cite{Aaij:2019jaq}. Among them, the isobar model is the most commonly used by BaBar, Belle and LHCb experiments. Based on the isobar approximation, the decay amplitude can be modeled as a coherent combination of all individual decay channels, which can be expressed as
\begin{eqnarray}\label{isobar}
\mathcal{A}=\sum_{i=1}^{N}C_i\mathcal{A}_i,
\end{eqnarray}
where the $\mathcal{A}_i$ is the decay amplitude of individual decay channel. The complex coefficients $C_i$ describes the relative magnitude and the phase of each decay channels. On the theoretical side, the three-body $B$ decays have been studied extensively in various methods, such as the approaches based on the symmetry principles \cite{Gronau:2005ax, Engelhard:2005hu, Imbeault:2011jz, Bhattacharya:2013boa, He:2014xha}, QCD factorization (QCDF) approach\cite{ElBennich:2009da, Krankl:2015fha, Cheng:2002qu, Cheng:2007si, Cheng:2016shb, Cheng:2014uga, Cheng:2013dua, Li:2014oca,Huber:2020pqb}, perturbative QCD approach (PQCD) \cite{Chen:2002th, Wang:2016rlo, Li:2016tpn, Ma:2019sjo, Li:2018qrm, Li:2018lbd, Ma:2017aie, Li:2017obb, Ma:2017kec, Li:2017mao, Ma:2017idu, Ma:2016csn, Li:2015tja, Zou:2020atb, Zou:2020fax, Zou:2020khk}, and other theoretical methods \cite{Zhang:2013oqa, Wang:2015ula, ElBennich:2006yi}.

In 2017, LHCb Collaboration reported the measurement of the decay $B_s^0\to \phi \pi^+\pi^-$ and the evidence of $B^0\to \phi\pi^+\pi^-$ \cite{Aaij:2016qnm}. Based on the combined analysis of the $\pi^+\pi^-$ mass spectrum and the angles among the final states, the corresponding branching fractions of the quasi-two-body decays are given as \footnote{For the convenience, we abbreviate $f_0(980)$ and $f_2(1270)$ as $f_0$ and $f_2$ in the following.}
\begin{eqnarray}
\mathcal{B}(B_s^0\to \phi (f_0(980)\to)\pi^+\pi^-)&=&(1.12\pm0.16^{+0.09}_{-0.08}\pm0.11)\times 10^{-6},\\
\mathcal{B}(B_s^0\to \phi (f_2(1270)\to)\pi^+\pi^-)&=&(0.61\pm0.13^{+0.12}_{-0.05}\pm0.06)\times 10^{-6}.
\end{eqnarray}
Stimulated by above results, we shall investigate above two quasi-two-body decays in the PQCD approach. As aforementioned, in the $B$ meson rest frame, the $\pi\pi$ pair and the bachelor $\phi$ meson move fast and back-to-back, which indicates that the interaction between the $\pi\pi$ pair and the bachelor $\phi$ meson is highly suppressed. On the other hand, the interactions between the two pions in the $\pi\pi$ pair can be described by a two-meson wave function, which includes both resonant and nonresonant contributions. It is obvious that these quasi-two-body decays are very similar to the two-body decays, by substituting one particle by a system involving two fast moving particles, the factorization formalism would be applicable.

In the PQCD approach that is based on $k_T$ factorization, the physics with the scale above the $W$ boson mass $m_W$ is weak interaction, which can be calculated perturbatively. Using the obtained Wilson coefficients at the scale $m_W$ and the renormalization group equation we can evaluate the Wilson coefficients including the physics from the $M_W$ to the $m_b$, the $b$ quark mass. In the PQCD picture, the physics between the scale $m_b$ and the factorization scale $\Lambda_h$ is regarded to be dominated by the hard gluon exchange, and can be perturbatively calculated, which is the so-called hard kernel $\mathcal{H}$. The soft dynamics below the factorization scale is nonperturbative, which can be described by the universal hadronic wave functions of the initial and final states. Therefore, the decay amplitude of the quasi-two-body $B_{(s)}^0 \to \phi (f_0 /f_2\to) \pi^+\pi^-$ decays can thus be factorized as the convolution
\begin{eqnarray}
\mathcal{A}=C(t)\otimes \mathcal{H}(x_i,b_i,t)\otimes \Phi_B(x_1,b_1)\otimes\Phi_{\phi}(x_2,b_2)\otimes\Phi_{\pi\pi}(x_3,b_3)\otimes\exp^{-S(t)},
\end{eqnarray}
where the $x_i$ are the momentum fraction of the quarks, $b_i$ are the conjugate variables of the quarks' transverse momenta $k_{iT}$, and $t$ is the largest scale appearing in the hard kernel $\mathcal{H}(x_i,b_i,t)$. $\Phi_B$ and $\Phi_{\phi}$ are the wave functions of the $B$ meson and $\phi$ meson, and the $\Phi_{\pi\pi}$ is the $\pi\pi$ pair wave function. The exponential term is the so-called Sudakov form factor caused by the additional scale introduced by the intrinsic transverse momenta $k_T$ of the quarks, which suppresses the soft dynamics effectively\cite{Li:2001ay,Li:1997un,Lu:2000hj}.

The paper is organized as follows: In Sec.~\ref{sec:function}, we  introduce the wave functions used in the PQCD calculations, and the theoretical decay amplitudes are also presented in this section. The numerical results and the discussions are given in Sec.~\ref{sec:result}. Finally, we summarize this work in Sec.\ref{summary}.
\section{Decay Formalism}\label{sec:function}
In SM, the effective Hamiltonian $\mathcal{H}_{eff}$ for the quark-level transition $b\to s q\bar q$ governing the considered quasi-two-body decays is given as \cite{Buchalla:1995vs}
 \begin{eqnarray}\label{effhamil}
\mathcal{H}_{eff}=\frac{G_F}{\sqrt{2}}\left\{V_{ub}V^*_{us}(C_1O_1+C_2O_2)-V_{tb}V^*_{ts}\sum_{i=3}^{10}C_i O_i\right\},
\end{eqnarray}
where $V_{IJ}$ are  the Cabibbo-Kobayashi-Maskawa (CKM) matrix elements and $G_F$ is the Fermi constant. The explicit expressions for the local four-quark operators $O_i$ ($i = 1, ..., 10$) and their corresponding Wilson coefficients $C_i$ can be found in Ref.~\cite{Buchalla:1995vs}.

In PQCD, the most important inputs are the the wave functions of the initial and final states, including the $B$ meson, the $\phi$ meson and the $\pi\pi$ pair. The wave functions of the $B$ meson and the $\phi$ meson have been well determined by the two-body $B$ decays \cite{Liu:2019ymi, Zou:2016yhb, Qin:2014xta, Yu:2013pua, Zou:2012sx, Li:2015xna, Wang:2017hxe, Zhou:2015jba, Yu:2005rh, Li:2004ep, Colangelo:2010wg, Colangelo:2010bg, Wang:2016wpc}, and the wave functions of $\pi\pi$ pair corresponding to the different spins have been also discussed in refs. \cite{Li:2018lbd, Ma:2017aie, Li:2017obb, Xing:2019xti, Meissner:2013hya}. In this work we will discuss the contributions of $S$-wave and $T$-wave $\pi\pi$ pair wave functions, which are related to the intermediate resonances $f_0$ and $f_2$, respectively. For the $S$-wave two-pion wave function, its structure is given as \cite{Wang:2018xux,Xing:2019xti}
\begin{eqnarray}
\Phi_{\pi\pi}^S=\frac{1}{\sqrt{2N_c}}\left[P\mkern-10.5mu/\phi_S(z,\xi,\omega)+\omega\phi_S^s(z,\xi,\omega)+\omega(n\mkern-10.5mu/v\mkern-10.5mu/-1)
\phi_S^t(z,\xi,\omega)\right],
\end{eqnarray}
where $z$ represents the momentum fraction of the light quark in the $\pi\pi$ pair, and $\xi$ is the momentum fraction of one pion in the $\pi\pi$ pair. The $P$ is the momentum of the $\pi\pi$ pair, satisfying the condition $P^2=\omega^2$ with the invariant mass of $\pi\pi$ pair. The $n$ and $v$ are the light-like vectors. The $\phi_S$ and $\phi_S^{s,t}$ are the twist-2 and twist-3 light-cone distribution amplitudes, which are given explicitly as \cite{Wang:2014ira,Wang:2015uea}
\begin{eqnarray}
\phi_S(z,\xi,\omega)&=&\frac{F_S(\omega)}{\sqrt{6}}9 a_sz(1-z)(2z-1), \\
\phi_S^s(z,\xi,\omega)&=&\frac{F_S(\omega)}{2\sqrt{6}},\\
\phi_S^t(z,\xi,\omega)&=&\frac{F_S(\omega)}{2\sqrt{6}}(1-2z),
\end{eqnarray}
with $F_S(\omega)$ being the time-like from factor. For a narrow intermediate resonance, $F_S(\omega)$ is particularly described successfully by the Breit-Wigner line-shape. However, for the resonance $f_0(980)$, due to the abnormal enhancement from the $KK$ system found around $980~ {\rm MeV}$ in the $\pi\pi$ scattering, the Breit-Wigner line-shape of $f_0(980)$ is modified to be Flatt$\acute{e}$ model \cite{Flatte:1976xu}. In our calculations, We here adopt the formulae updated by LHCb collaboration in ref.\cite{Aaij:2014emv},
\begin{eqnarray}
F_S(\omega)=\frac{m_{f_0}^2}{m^2_{f_0}-\omega^2-i m_{f_0}(g_{\pi\pi}\rho_{\pi\pi}+g_{KK}\rho_{KK}F^2_{KK})},
\end{eqnarray}
where
\begin{eqnarray}
\rho_{\pi\pi}&=\sqrt{1-\frac{4m^2_{\pi}}{\omega^2}},\,\,\,\rho_{KK}&=\sqrt{1-\frac{4m_K^2}{\omega^2}}.
\end{eqnarray}
In above functions, $g_{\pi\pi}$ and $g_{KK}$ are the coupling constants extracted from $f_0(980)\to \pi\pi$ and $f_0(980)\to KK$ decays, respectively, whose values are taken as $g_{\pi\pi}=(0.165\pm0.018)~{\rm GeV}^2$ and $g_{KK}/g_{\pi\pi}=4.21\pm0.33$. The factor $F_{KK}= e^{-\alpha q^2}$ with $\alpha=-2.0$ could suppress the contribution from $KK$ scattering. In this work we shall use the Gegenbauer moment $a_s=0.3\pm0.2$, which is in agreement with that determined in ref.\cite{Xing:2019xti}.

Now, we turn to discuss the $D$-wave $\pi\pi$ pair wave function. Because of the conservation of angular momentum, the $\pm2$ polarization components of tensor structure can not contribute to the concerned decay amplitudes. So, the behavior of the tensor structure in $B$ meson decays is very similar to the vector one. In order to describe the contribution of tensor structure conveniently, we can define a new polarization vector $\epsilon^{\prime}$ associated with the polarization tensor $\epsilon_{\mu\nu}$ of tensor state. As the treatments in refs.\cite{Qin:2014xta, Kim:2013cpa, Zou:2012xk, Zou:2012sy, Zou:2012sx, Zou:2012td, Wang:2010ni, Cheng:2010yd}, the new defined vector is proportional to the polarization vector of the vector mesons with the coefficients $\sqrt{\frac{2}{3}}$ and $\sqrt{\frac{1}{2}}$ for the longitudinal and transverse polarizations, respectively.  Therefore, for the $D$-wave two-pion system, the longitudinal and transverse wave functions are expressed as \cite{Li:2018lbd}
\begin{eqnarray}
&&\Phi_D^L(\pi\pi)=\frac{1}{\sqrt{2N_c}}\left[P\mkern-10.5mu/\phi_D(z,\xi,\omega)+\omega\phi_D^s(z,\xi,\omega)+\frac{P\mkern-10.5mu/_1 P\mkern-10.5mu/_2-P\mkern-10.5mu/_2P\mkern-10.5mu/_1}{\omega(2\xi-1)}\phi_D^t(z,\xi,\omega)\right],\\
\label{Dwave-L}
&&\Phi_D^T(\pi\pi)=\frac{1}{\sqrt{2N_c}}\left[\gamma_5 \epsilon^{\prime}_T\mkern-15.5mu/\; P\mkern-9.5mu/\phi_D^T(z,\xi,\omega)+\omega\gamma_5\epsilon^{\prime}_T\mkern-15.5mu/\;\phi_D^a(z,\xi,\omega)+i\omega
\frac{\epsilon^{\mu\nu\rho\sigma}\gamma_{\mu}\epsilon^{\prime}_{T\nu}P_{\rho}v_{\sigma}}{P\cdot v}\phi_D^v(z,\xi,\omega)\right],
\end{eqnarray}
where $\phi_D(z,\zeta,\omega)$ and $\phi_D^{T}(z,\zeta,\omega)$ are the twist-2 light-cone distribution amplitudes associating to the longitudinal and transverse polarization components respectively, and the rest $\phi_D^{s,t}(z,\zeta,\omega)$  and $\phi_D^{a,v}(z,\zeta,\omega)$ are the twist-3 light-cone distribution amplitudes. The explicit expressions of the twist-2 distribution amplitudes are the same as those of $KK$ pair discussed in refs.\cite{Zou:2020atb,Rui:2019yxx}
\begin{eqnarray}
\phi_D(z,\xi,\omega)&=&\sqrt{\frac{2}{3}}\frac{9F_D^{\parallel}(\omega)}{\sqrt{2N_c}}a_D z(1-z)(2z-1)\zeta(\xi),\\
\phi_D^T(z,\xi,\omega)&=&\sqrt{\frac{1}{2}}\frac{9F_D^{\perp}(\omega)}{\sqrt{2N_c}}a_D^Tz(1-z)(2z-1)\tau(\xi),
\end{eqnarray}
with the factors $\zeta(\xi)$ and $\tau(\xi)$ describing the phase space of two-pion pair
\begin{eqnarray}
\zeta(\xi)=1-6\xi+6\xi^2,\;\;\tau(\xi)=(2\xi-1)\sqrt{\xi(1-\xi)}.
\end{eqnarray}
$F_D^{\parallel,\perp}$ are the $D$-wave two-pion time-like form factor that can be well modeled by the Breit-Wigner line shape, and the detailed forms can be given as \cite{Zyla:2020zbs}:
\begin{eqnarray}
F_D^{\parallel}(\omega)=\frac{m_R^2}{m_R^2-\omega^2-im_R\Gamma(\omega)},
\end{eqnarray}
with the nominal mass $m_R$ being the mass of the parent resonance. The dependence of the decay width of the resonance on invariant mass $\omega$ is given by
\begin{eqnarray}
\Gamma(\omega)=\Gamma_0\left(\frac{|q|}{|q_0|}\right)^5\frac{m_R}{\omega}X^2(\kappa),
\end{eqnarray}
with the nominal decay width of resonance $\Gamma_0$. Moreover, $|q_0|$ is the value of daughter $\pi$'s momentum $|q|$ when $\omega=m_R$. The factor $X(\kappa)$ is the Blatt-Weisskopf angular momentum barrier factor\cite{Blatt:1952ije}, whose expression can be given as
\begin{eqnarray}
X(\kappa)=\sqrt{\frac{9+3\kappa_0^2+\kappa_0^4}{9+3\kappa^2+\kappa^4}}\,\,,\kappa=r|q|.
\end{eqnarray}
The effective radius $r$ of the intermediate resonance cannot affect the numerical results remarkably and is chosen to be $r=4.0\,{\rm GeV}^{-1}$, following the experimental analysis\cite{Lees:2012kxa}. $\kappa_0$ is the value of the $\kappa$ when $\omega=m_R$. As for the transverse time-like form factor $F_D^{\perp}$, it can be determined by the approximate relation \cite{Wang:2016rlo}
\begin{eqnarray}
\frac{F_D^{\perp}}{F_D^{\parallel}}\simeq\frac{f_R^{T}}{f_R},
\end{eqnarray}
with $f_R^{(T)}$ being the (transvers) decay constant of the tensor resonance. For the rest twist-3 light-cone distribution amplitudes, the expressions are given by
\begin{eqnarray}
\phi_D^s(z,\xi,\omega)&=&\sqrt{\frac{2}{3}}\frac{-9F_D^{\perp}(\omega)}{4\sqrt{2N_c}}a_D(1-6z+6z^2)\zeta(\xi),\nonumber\\
\phi_D^t(z,\xi,\omega)&=&\sqrt{\frac{2}{3}}\frac{9F_D^{\perp}(\omega)}{4\sqrt{2N_c}}(2z-1)(1-6z+6z^2)\zeta(\xi),\nonumber\\
\phi_D^a(z,\xi,\omega)&=&\sqrt{\frac{1}{2}}\frac{3F_D^{\parallel}}{2\sqrt{2N_c}}a_D^T(2z-1)^3\tau(\xi),\nonumber\\
\phi_D^v(z,\xi,\omega)&=&\sqrt{\frac{1}{2}}\frac{-3F_D^{\parallel}}{2\sqrt{2N_c}}a_D^T(1-6z+6z^2)\tau(\xi).
\end{eqnarray}
In ref.\cite{Li:2018lbd}, the authors discussed the processes $B_{(s)}\to Pf_2\to P\pi\pi$ decays with $f_2$ as the intermediate resonance, $P$ being a pseudoscalar meson, and the longitudinal Gegenbauer moment $a_D=0.4\pm0.1$ had been determined. In addition, $a_D^T=0.8\pm0.2$ can also be fixed from experimental results of $B\to K\pi\pi$ decays \cite{Garmash:2006fh}.

\begin{figure}[!htb]
\begin{center}
\includegraphics[scale=0.5]{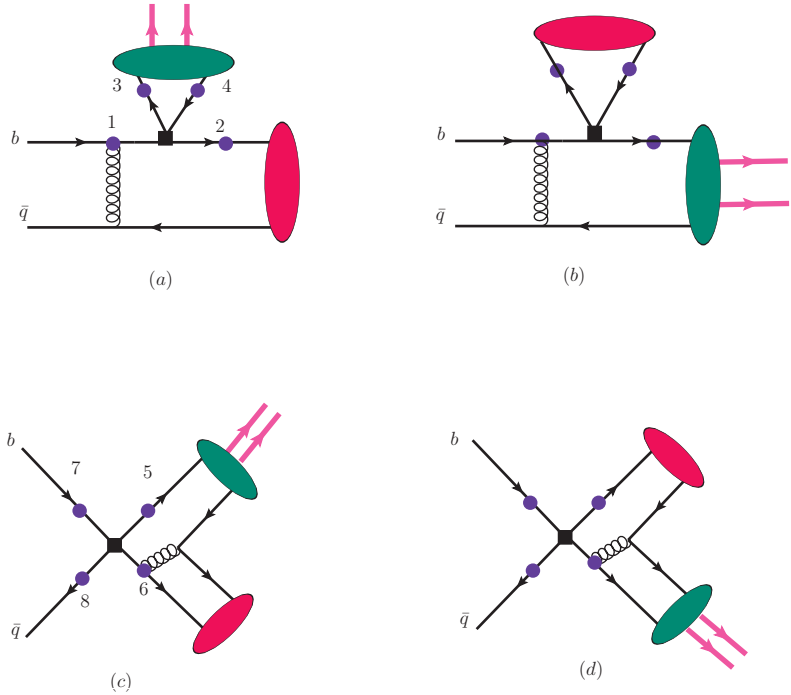}
\caption{Typical Feynman diagrams for the quasi-two-body decay $B_{(s)}\to \phi \pi\pi$ in PQCD, where the black squares stand for the weak vertices, and large (purple) spots on the quark lines denote possible attachments of hard gluons. The green ellipse represent $\pi\pi$-pair and the red one is the light bachelor $\phi$ meson.}\label{feynman}
\end{center}
\end{figure}

Based on the effective Hamiltonian and the wave functions, we can perform the theoretical calculation in the PQCD approach. In the leading order, the diagrams contributing to the decay amplitude are plotted in the Figure.\ref{feynman}.  The first two diagrams are the emission type diagrams, where diagram (a) is the $\pi\pi$ emission and the $\phi$ meson is emitted in diagram (b).  The last two diagrams are the annihilation diagrams, where the produced antiquark flows into the $\phi$ meson in diagram (c) and flows into $\pi\pi$-pair in diagram (d).

We take $B_{s,d}\to \phi (f_0\to)\pi^+\pi^- $ decays as examples for illustration. At first, we discuss the contributions of the emission diagrams, as shown in Figures.~\ref{feynman} (a) and (b). If the hard gluons are from the spots ``1" and ``2", the $\pi\pi$ pair or $\phi$ can be factorized out, and we call them factorizable diagrams. When inserting the $(V-A)(V-A)$ and $(V-A)(V+A)$ current, due to the charge conjugation invariance, the $S$-wave two-pion pair cannot be emitted, namely
\begin{eqnarray}
F_{\pi\pi}^{LL}=F_{\pi\pi}^{LR}=0 \label{pipi},
\end{eqnarray}
where the subscript $\pi\pi$ means two-pion pair is emitted, and the superscripts $LL$ and $LR$ indicate the inserted $(V-A)(V-A)$ and $(V-A)(V+A)$ currents, respectively. When the $\phi$ meson is emitted, the amplitude is given by
\begin{eqnarray}
F_{\phi}^{LL,LR}&=&8\pi C_F f_V m_B^4 \int_{0}^{1}dx_1dz\int_0^{\infty}b_1db_1b_zdb_z \phi_B(x_1,b_1)\frac{1}{\sqrt{1-\eta^2}}\nonumber\\
&&\Big\{\Big[(\eta^2-1)(z+1)\phi_S(z,\xi,\omega)+\eta(2z-1)(\phi_S^s(z,\xi,\omega)+\phi_S^t(z,\xi,\omega))\Big]\nonumber\\
&&\left.\times E_{ef}(t_a)h_{ef}(x_1,z(1-r_V^2),b_1,b_z)\right.\nonumber\\
&& +\eta\Big[\phi_S(z,\xi,\omega)-2\phi_S^s(z,\xi,\omega)\Big] E_{ef}(t_b)h_{ef}(z,x_1(1-r_V^2),b_z,b_1)\Big\}.
\end{eqnarray}
with $r_V=m_{\phi}/m_B$ and $\eta=\omega/m_B$. In the above formular, $C_F=\frac{4}{3}$ is the color factor and $f_V$ is the decay constant of the $\phi$ meson. The functions $E_{ef}$, $h_{ef}$ and the typical scales $t_{a,b}$ are referred to ref.\cite{Zou:2015iwa}. If we insert the $(S-P)(S+P)$ current arising from the Fierz transformation of the $(V-A)(V+A)$ current, the contribution from diagram with two-pion pair emission is given by
\begin{eqnarray}
F_{\pi\pi}^{SP}&=&-16\pi C_F F_S \eta m_B^4\int_0^1dx_1dx_3\int_0^{\infty}b_1db_1b_3db_3\phi_B(x_1,b_1)\Big\{\Big[\phi_V^s(x_3)r_V(x_3+2)\nonumber\\
&&\left.+\sqrt{1-\eta^2}(\phi_V(x_3)-\phi_V^t(x_3)r_Vx_3)\Big]E_{ef}(t_a)h_{ef}(x_1,x_3(1-\eta^2),b_1,b_3)\right.\nonumber\\
&&\left.+2r_V\eta\phi_V^s(x_3)E_{ef}^{t_b}h_{ef}(x_3,x_1(1-\eta^2),b_3,b_1)\right\},
\end{eqnarray}
with the $S$ wave  time-like form factor $F_S$ of the $\pi\pi$ pair. For the diagrams with the emitted vector $\phi$ meson, the decay amplitude vanishes due to the fact that the vector meson can not be produced through the $(S\pm P)$ currents \cite{Ali:2007ff}, i.e.
\begin{eqnarray}
F_{\phi}^{SP}=0.
\end{eqnarray}

If the hard gluons come from spots ``3" and ``4", all three wave functions are involved, and we call them the nonfactorizable emission diagrams. Their amplitudes with different currents can be written as
\begin{eqnarray}
M_{\pi\pi}^{LL}&=&16\sqrt{\frac{2}{3}}\pi C_F m_B^4\int_0^1dx_1dx_3dz\int_0^{\infty}b_1db_1b_3db_3b_zdb_z\phi_B(x_1,b_1)\phi_S(z,\xi,\omega)\nonumber\\
&&\Big\{\Big[\sqrt{1-\eta^2}(\phi_V(x_3)(\eta^2+1)(z-1)-\phi_V^t(x_3)r_Vx_3)+\phi_V^s(x_3)r_Vx_3\Big]\nonumber\\
&&\times E_{enf}(t_c)h_{enf}(\alpha,\beta_1,b_1,b_z)\nonumber\\
&&+\Big[\sqrt{1-\eta^2}((z+x_3)\phi_V(x_3)-r_Vx_3\phi_V^t(x_3))-r_Vx_3\phi_V^s(x_3)\Big]\nonumber\\
&&\times E_{enf}(t_d)h_{enf}(\alpha,\beta_2,b_1,b_z)\Big\},
\\
M_{\phi}^{LL}&=&-16\sqrt{\frac{2}{3}}\pi C_F m_B^4\int_0^1dx_1dzdx_3\int_0^{\infty}b_1db_1b_3db_3\phi_B(x_1,b_1)\phi_V(x_3)\sqrt{1-\eta^2}\nonumber\\
&&\Big\{\Big[\phi(z,\xi,\omega)(1-x_3)+z\eta(\phi_S^t(z,\xi,\omega)-\phi_S^s(z,\xi,\omega))\Big]
E_{enf}(t_c)h_{enf}(\alpha,\beta_1,b_1,b_3) \nonumber\\
&& +\Big[z\eta(\phi_S^s(z,\xi,\omega)+\phi_S^t(z,\xi,\omega))-\phi_S(z,\xi,\omega)(z+x_3)\Big]
E_{enf}(t_d)h_{enf}(\alpha,\beta_2,b_1,b_3)\Big\},
\\
M_{\pi\pi}^{LR}&=&16\sqrt{\frac{2}{3}}\pi C_F \eta m_B^4\int_0^1dx_1dzdx_3\int_0^{\infty}b_1db_1b_zdb_z\phi_B(x_1,b_1)\nonumber\\
&&\left\{\left[\phi_S^s(z,\xi,\omega)(\sqrt{1-\eta^2}(\phi_V(x_3)(1-z)+\phi_V^t(x_3)r_V(x_3+z-1))+\phi_V^s(x_3)r_V(x_3-z+1))\right.\right.\nonumber\\
&&\left.\left.-\phi_S^t(z,\xi,\omega)(\sqrt{1-\eta^2}(\phi_V(x_3)(z-1)+\phi_V^t(x_3)r_V(x_3-z+1))+\phi_V^s(x_3)r_v(x_3+z-1))\right]\right.\nonumber\\
&&\left.\times E_{enf}(t_c)h_{enf}(\alpha,\beta_1,b_1,b_z)\right.\nonumber\\
&&\left.-\left[\phi_S^s(z,\xi,\omega)(\sqrt{1-\eta^2}(\phi_V(x_3)z+r_V\phi_V^t(x_3)(x_3-z))+r_V\phi_V^s(x_3)(x_3+z))\right.\right.\nonumber\\
&&\left.\left.+\phi_S^t(z,\xi,\omega)(\sqrt{1-\eta^2}(r_V\phi_V^t(x_3)(x_3+z)-z\phi_V(x_3))+r_V\phi_V^s(x_3)(x_3-z))\right]\right.\nonumber\\
&& \times E_{enf}(t_d)h_{enf}(\alpha,\beta_2,b_1,b_z)\Big\},
\\
M_{\phi}^{LR}&=&16\sqrt{\frac{2}{3}}\pi C_F r_V M_B^4\int_0^1dx_1dx_3dz\int_0^{\infty}b_1db1b_3db3\phi_B(x_1,b_1)\nonumber\\
&&\left\{\left[\phi_S(z,\xi,\omega)(\phi_V^s(x_3)+\phi_V^t(x_3)\sqrt{1-\eta^2})(1-x_3)\right.\right.\nonumber\\
&&\left.\left.+\eta\left(\phi_S^s(z,\xi,\omega)(\phi_V^s(x_3)(z-x_3+1)-\phi_V^t(x_3)\sqrt{1-\eta^2}(z+x_3-1))\right.\right.\right.\nonumber\\
&&\left.\left.\left.+\phi_S^t(z,\xi,\omega)(\phi_V^s(x_3)(z+x_3-1)+\phi_V^t(x_3)\sqrt{1-\eta^2}(x_3-z-1))\right)\right]\right.\nonumber\\
&&\left.\times E_{enf}(t_c)h_{enf}(\alpha,\beta_1,b_1,b_3)\right.\nonumber\\
&&\left.-\left[\phi_S(z,\xi,\omega)(\phi_V^s(x_3)x_3-x_3\sqrt{1-\eta^2}\phi_V^t(x_3))\right.\right.\nonumber\\
&&\left.\left.+\eta\left(\phi_S^s(z,\xi,\omega)(\phi_V^s(x_3)(z+x_3)+\phi_V^t(x_3)\sqrt{1-\eta^2}(z-x_3))\right.\right.\right.\nonumber\\
&&\left.\left.\left.+\phi_S^t(z,\xi,\omega)(\phi_V^s(x_3)(z-x_3)+\phi_V^t(x_3)\sqrt{1-\eta^2}(z+x_3))\right)\right]\right.\nonumber\\
&&\times E_{enf}(t_d)h_{enf}(\alpha,\beta_2,b_1,b_3)\Big\},
\\
M_{\pi\pi}^{SP}&=&16\sqrt{\frac{2}{3}}\pi C_F m_B^4\int_0^1dx_1dzdx_3\int_0^{\infty}b_1db_1b_zdb_z\phi_B(x_1,b_1)\phi_S(z\xi,\omega) \nonumber\\
&&\Big\{\Big[r_Vx_3\phi_V^s(x_3)+\sqrt{1-\eta^2}(r_Vx_3\phi_V^t(x_3)-\phi_V(x_3)((1-\eta^2)x_3+1-z+\eta^2)\Big]\nonumber\\
&&\left.\times E_{enf}(t_c)h_{enf}(\alpha,\beta_1,b_1,b_z)\right.\nonumber\\
&&\left.-\left[r_Vx_3\phi_V^s(x_3)-\sqrt{1-\eta^2}(r_Vx_3\phi_V^t(x_3)+\phi_V(x_3)(1-\eta^2)z)\right]\right.\nonumber\\
&&\times E_{enf}(t_d)h_{enf}(\alpha,\beta_2,b_1,b_z)\Big\},
\\
M_{\phi}^{SP}&=&16\sqrt{\frac{2}{3}}\pi C_F m_B^4\sqrt{1-\eta^2}\int_0^1dx_1dx_3dz\int_0^{\infty}b_1db_1b_3db_3\phi_B(x_1b_1)\phi_V(x_3)\nonumber\\
&&\Big\{\Big[\phi_S(z,\xi,\omega)(z-(1-\eta^2)(x_3-1))-z\eta(\phi_S^s(z,\xi,\omega)+\phi_S^t(z,\xi,\omega))\Big]\nonumber\\
&&\times E_{enf}(t_c)h_{enf}(\alpha,\beta_1,b_1,b_3)\nonumber\\
&&-\Big[\phi_S(z,\xi,\omega)(x_3+\eta^2(z-x_3))+z\eta(\phi_S^s(z,\xi,\omega)+\phi_S^t(z,\xi,\omega))\Big]\nonumber\\
&&\times E_{enf}(t_d)h_{enf}(\alpha,\beta_2,b_1,b_3)\Big\}.
\end{eqnarray}
All functions $E_{enf}$, $h_{enf}$, $\alpha$ and $\beta_{1,2}$ can also be found in ref.\cite{Zou:2015iwa}.

Now, let us deal with the annihilation diagrams. Unlike QCD factorization approach where the annihilation diagrams cannot be calculated reliably, in the framework of PQCD the annihilation diagrams can be calculated without endpoint singularity by keeping the intrinsic transverse momentum of each quark. Similarly, the annihilation diagrams can also be divided into two kinds, factorizable and nonfactorizable. If the gluons come from the spots ``5" and ``6", the wave function of $B$ meson can be factorized out, and the corresponding diagrams are called factorizable annihilation diagrams. If the gluons are emitted from the spots ``7" and ``8", their contributions are nonfactorizable. For the factorizable annihilation diagrams, the amplitudes associated with the $(V-A)(V-A)$,$(V-A)(V+A)$ and $(S-P)(S+P)$ currents are
\begin{eqnarray}
A_{\pi\pi}^{LL}&=&8\pi C_Ff_Bm_B^4\int_0^1dzdx_3\int_0^{\infty}b_zdb_zb_3db_3\Big\{\Big[\phi_S(z,\xi,\omega)\phi_V(x_3)\sqrt{1-\eta^2}(1+x_3(\eta^2-1))
\nonumber\\
&&+2\phi_S^s(z,\xi,\omega)r_V\eta(\phi_V^s(x_3)(x_2-2)-\phi_V^t(x_3)x_3\sqrt{1-\eta^2})\Big]\  E_{af}(t_e)h_{af}(\alpha_1,\beta,b_z,b_3)\nonumber\\
&&-\Big[\phi_S(z,\xi,\omega)z\sqrt{1-\eta^2}-2\phi_V^s(x_3)r_V\eta(\phi_S^s(z,\xi,\omega)(z+1)
+\phi_S^t(z,\xi,\omega)(z-1))\Big]\nonumber\\
&& \times E_{af}(t_f)h_{af}(\alpha_2,\beta,b_3,b_z)\Big\},
\\
A_{\phi}^{LL}&=&8\pi C_Ff_Bm_B^4\int_0^1dx_3dz\int_0^{\infty}b_3db_3b_zdb_z\Big\{\Big[\phi_S(z,\xi,\omega)\phi_V(x_3)(1-z)\sqrt{1-\eta^2}\Big]\nonumber\\
&& +2r_V\eta\phi_V^s(x_3)(\phi_S^s(z,\xi,\omega)(z-2)-\phi_S^t(z,\xi,\omega)z)\Big]
 E_{af}(t_e)h_{af}(\alpha_1,\beta,b_3,b_z)\nonumber\\
&& -\Big[\phi_S(z,\xi,\omega)\phi_V(x_3)x_3-2r_V\eta\phi_S^s(z,\xi,\omega)(\phi_V^s(x_3)(1+x_3)+\phi_V^t(x_3)(x_3-1))\Big]\nonumber\\
&&\times E_{af}(t_f)h_{af}(\alpha_2,\beta,b_z,b_3)\Big\},
\\
A_{\pi\pi}^{LR}&=&A_{\pi\pi}^{LL},
\\
A_{\phi}^{LR}&=&A_{\phi}^{LL},
\\
A_{\pi\pi}^{SP}&=&16\pi C_Ff_Bm_B^4\int_0^1dzdx_3\int_0^{\infty}b_zdb_zb_3db_3\nonumber\\
&& \Big\{\Big[2\phi_S^s(z,\xi,\omega)\phi_V(x_3)\eta-r_V\phi_S(z,\xi,\omega)(\phi_V^s(x_3)+\phi_V^t(x_3))(1-x_3)\Big] E_{af}(t_e)h_{af}(\alpha_1,\beta,b_z,b_3) \nonumber\\
&& -\Big[2\phi_S(z,\xi,\omega)\phi_V^s(x_3)r_V+\phi_V(x_3)\eta z(\phi_S^t(z,\xi,\omega)-\phi_S^s(z,\xi,\omega))\Big] E_{af}(t_f)h_{af}(\alpha_2,\beta,b_3,b_z)\Big\},\nonumber\\
\\
A_{\phi}^{SP}&=&16\pi C_Ff_Bm_B^4\int_0^1dx_3dz\int_0^{\infty}b_3db_3b_zdb_z\nonumber\\
&&\Big\{\Big[2r_V\phi_S(z,\xi,\omega)\phi_V^s(x_3)-(1-z)\eta\phi_V(x_3)(\phi_S^s(z,\xi,\omega)+\phi_S^t(z,\xi,\omega))\Big]\nonumber\\
&&\times E_{af}(t_e)h_{af}(\alpha_1,\beta,b_3,b_z)  -\Big[2\eta\phi_V(x_3)\phi_S^s(z,\xi,\omega)-r_Vx_3\phi_S(z,\xi,\omega)(\phi_V^s(x_3)+\phi_V^t(x_3))\Big]\nonumber\\
&&\times E_{af}(t_f)h_{af}(\alpha_2,\beta,b_z,b_3)\Big\}.
\end{eqnarray}
Within the same way, we obtain the expressions of the amplitudes of the nonfactorizable annihilation diagrams as
\begin{eqnarray}
W_{\pi\pi}^{LL}&=&16\sqrt{\frac{2}{3}}\pi C_Fm_B^4\int_0^1dx_1dzdx_3\int_0^{\infty}b_1db_1b_zdb_z\phi_B(x_1,b_1)\Big\{\Big[z\phi_S(z,\xi,\omega)\phi_V(x_3)\sqrt{1-\eta^2}\nonumber\\
&&+r_V\eta\Big(\phi_S^s(z,\xi,\omega)(\phi_V^s(x_3)(x_3-z-3)+\phi_V^t(x_3)(x_3+z-1)) \nonumber\\
&& -\phi_S^t(z,\xi,\omega)(\phi_V^s(x_3)(x_3+z-1)+\phi_V^t(x_3)(x_3-z+1))\Big)\times E_{anf}(t_g)h_{anf}(\alpha,\beta_1,b_1,b_z) \nonumber\\
&&-\Big[\phi_S(z,\xi,\omega)\phi_V(x_3)(1-x_3)+r_V\eta\Big(\phi_S^s(x_3)(\phi_V^s(x_3)(x_3-z-1)-\phi_V^t(x_3)(x_3+z-1))\nonumber\\
&&+\phi_S^t(z,\xi,\omega)(\phi_V^s(x_3)(x_3+z-1)+\phi_V^t(x_3)(1-x_3+z))\Big)\Big]E_{anf}(t_h)h_{anf}(\alpha,\beta_2,b_1,b_z)\Big\},
\\
W_{\phi}^{LL}&=&16\sqrt{\frac{2}{3}}\pi C_Fm_B^4\int_0^1dx_1dx_3dz\int_0^{\infty}b_1db_1b_3db_3\phi_B(x_1,b_1)\Big\{\Big[x_3\phi_S(z,\xi,\omega)\phi_V(x_3)\sqrt{1-\eta^2}\nonumber\\
&&+r_V\eta\Big(\phi_S^s(z,\xi,\omega)(\phi_V^s(x_3)(z-x_3-3)-\phi_V^t(x_3)(z+x_3-1))\nonumber\\
&&+\phi_S^t(z,\xi,\omega)(\phi_V^s(x_3)(z+x_3-1)+\phi_V^t(x_3)(x_3-z-1))\Big) E_{anf}(t_g)h_{anf}(\alpha,\beta_1,b_1,b_3)\nonumber\\
&&-\Big[\phi_S(z,\xi,\omega)\phi_V(x_3)\sqrt{1-\eta^2}(1-z)+r_V\eta\Big(\phi_S^s(z,\xi,\omega)(\phi_V^s(x_3)(z-x_3-1)\nonumber\\
&&+\phi_V^t(x_3)(z+x_3-1))+\phi_S^t(z,\xi,\omega)(\phi_V^t(x_3)(1-z+x_3)-\phi_V^s(x_3)(z+x_3-1))\Big)\Big]\nonumber\\
&&\times E_{anf}(t_h)h_{anf}(\alpha,\beta_2,b_1,b_3)\Big\},
\\
W_{\pi\pi}^{LR}&=&16\sqrt{\frac{2}{3}}\pi C_Fm_B^4\int_0^1dx_1dzdx_3\int_0^{\infty}b_1db_1b_zdb_z\phi_B(x_1,b_1)\nonumber\\
&&\Big\{\Big[r_V\phi_S(z,\xi,\omega)(\phi_V^s(x_3)-\phi_V^t(x_3))(1+x_3)
+\phi_V(x_3)\eta(2-z)(\phi_S^s(z,\xi,\omega)+\phi_S^t(z,\xi,\omega))\Big]\nonumber\\
&&\times E_{anf}(t_g)h_{anf}(\alpha,\beta_1,b_1,b_z)\nonumber\\
&&+\Big[r_V\phi_S(z,\xi,\omega)(\phi_V^t(x_3)-\phi_V^s(x_3))(x_3-1)+\phi_V(x_3)\eta z(\phi_S^s(z,\xi,\omega)+\phi_S^t(z,\xi,\omega))\Big]\nonumber\\
&& \times E_{anf}(t_h)h_{anf}(\alpha,\beta_2,b_1,b_z)\Big\},
\\
W_{\phi}^{LR}&=&16\sqrt{\frac{2}{3}}\pi C_Fm_B^4\int_0^1dx_1dx_3dz\int_0^{\infty}b_1db_1b_3db_3\phi_B(x_1,b_1)\nonumber\\
&&\Big\{\Big[r_V\phi_S(z,\xi,\omega)(\phi_V^s(x_3)+\phi_V^t(x_3))(2-x_3)
+\eta(1+z)\phi_V(x_3)(\phi_S^s(z,\xi,\omega)+\phi_S^t(z,\xi,\omega))\Big]\nonumber\\
&&\times E_{anf}(t_g)h_{anf}(\alpha,\beta_1,b_1,b_3)\nonumber\\
&&+\Big[r_Vx_3\phi_S(z,\xi,\omega)(\phi_V^t(x_3)+\phi_V^s(x_3))
+\eta(1-z)\phi_V(x_3)(\phi_S^s(z,\xi,\omega)-\phi_S^t(z,\xi,\omega))\Big]\nonumber\\
&& \times E_{anf}(t_h)h_{anf}(\alpha,\beta_2,b_1,b_3)\Big\},
\\
W_{\pi\pi}^{SP}&=&16\sqrt{\frac{2}{3}}\pi C_F m_B^4\int_0^1dx_1dzdx_3\int_0^{\infty}b_1db_1b_zdb_z\phi_B(x_1,b_1)\nonumber\\
&&\Big\{\Big[\phi_S(z,\xi,\omega)\phi_V(x_3)\sqrt{1-\eta^2}(1-x_3)+r_V\eta\Big(\phi_S^s(z,\xi,\omega)(\phi_V^s(x_3)(x_3-z-3)\nonumber\\
&&-\phi_V^t(x_3)(x_3+z-1)+\phi_S^t(z,\xi,\omega)(\phi_V^s(x_3)(x_3+z-1)+\phi_V^t(x_3)(z-x_3-1))\Big)\Big]\nonumber\\
&&\times E_{anf}(t_g)h_{anf}(\alpha,\beta_1,b_1,b_z)\nonumber\\
&&-\Big[\phi_S(z,\xi,\omega)\phi_V(x_3)\sqrt{1-\eta^2}z+r_V\eta\Big(\phi_S^s(z,\xi,\omega)(\phi_V^s(x_3)(x_3-z-1)\nonumber\\
&&+\phi_V^t(x_3)(x_3+z-1))+\phi_S^t(z,\xi,\omega)(\phi_V^t(x_3)(1-x_3+z)-\phi_V^s(x_3)(x_3+z-1))\Big)\Big]\nonumber\\
&&\times E_{anf}(t_h)h_{anf}(\alpha,\beta_2,b_1,b_z)\Big\},
\\
W_{\phi}^{SP}&=&16\sqrt{\frac{2}{3}}\pi C_F m_B^4\int_0^1dx_1dx_3dz\int_0^{\infty}b_1db_1b_3db_3\phi_B(x_1,b_1)\nonumber\\
&&\Big\{\Big[\phi_S(z,\xi,\omega)\phi_V(x_3)\sqrt{1-\eta^2}(1-z)+r_V\eta\Big(\phi_S^s(z,\xi,\omega)(\phi_V^s(x_3)(z-x_3-3)\nonumber\\
&&+\phi_V^t(x_3)(z+x_3-1))-\phi_S^t(z,\xi,\omega)(\phi_V^s(x_3)(z+x_3-1)+\phi_V^t(x_3)(z-x_3+1))\Big)\Big]\nonumber\\
&&\times E_{anf}(t_g)h_{anf}(\alpha,\beta_1,b_1,b_3)\nonumber\\
&&-\Big[\phi_S(z,\xi,\omega)\phi_V(x_3)\sqrt{1-\eta^2}x_3+r_V\eta\Big(\phi_S^s(z,\xi,\omega)(\phi_V^s(x_3)(z-x_3-1)\nonumber\\
&&-\phi_V^t(x_3)(z+x_3-1))+\phi_S^t(z,\xi,\omega)(\phi_V^s(x_3)(z+x_3-1)+\phi_V^t(x_3)(1-z+x_3))\Big)\Big]\nonumber\\
&&\times E_{anf}(t_h)h_{anf}(\alpha,\beta_2,b_1,b_3)\Big\},
\end{eqnarray}
where all related functions can also be found in ref.\cite{Zou:2015iwa}.

Within the obtained amplitudes with respect to the various currents, we can calculate the total decay amplitudes of the considered $B_{d,s}\to \phi (f_0\to)\pi^+\pi^-$ decays with the CKM matrix elements and the corresponding Wilson coefficients. It should be emphasized that for the quark structure of $f_0$ we adopt the two-quark picture with the mixing between the $q\bar{q} = \frac{1}{\sqrt{2}} (u\bar{u}+d\bar{d})$ and $s\bar{s}$, though the four-quark picture is also supported by some experimental results \cite{Aitala:2000xt}. As a result, the total decay amplitudes of the $B_{d,s}\to \phi (f_0\to)\pi^+\pi^-$ decays can be obtained as
\begin{eqnarray}
&&\mathcal{A}(B_{d}\to \phi (f_0\to)\pi^+\pi^-)=\mathcal{A}_{B_d}(q\bar{q})\sin\theta+\mathcal{A}_{B_d}(s\bar{s})\cos\theta,\\
&&\mathcal{A}(B_{s}\to \phi (f_0\to)\pi^+\pi^-)=\mathcal{A}_{B_s}(q\bar{q})\sin\theta+\mathcal{A}_{B_s}(s\bar{s})\cos\theta,
\end{eqnarray}
where $\mathcal{A}_{B_{d,s}}(q\bar{q})$ and $\mathcal{A}_{B_{d,s}}(s\bar{s})$ are the amplitudes from the $q\bar{q}$ and $s\bar{s}$ components respectively with the explicit expressions:
\begin{eqnarray}
\mathcal{A}_{B}(q\bar{q})&=&-\frac{G_F}{2}V_{td}V_{tb}^*\left\{\left(a_3-a_5+\frac{1}{2}(a_7-a_9)\right)F_{\phi}^{LL}
+\left(C_4-\frac{1}{2}C_{10}\right)M_{\phi}^{LL}\right.\nonumber\\
&&\left.+\left(C_6-\frac{1}{2}C_8\right)M_{\phi}^{SP}\right\},
\\
\mathcal{A}_B(s\bar{s})&=&-\frac{G_F}{\sqrt{2}}V_{td}V_{tb}^*\left\{\left(a_3+a_5-\frac{1}{2}(a_7-a_9)\right)(A_{\phi}^{LL}
+A_{\pi\pi}^{LL})\right.\nonumber\\
&&+\left(C_4-\frac{1}{2}C_{10}\right)(M_{\phi}^{LL}+M_{\pi\pi}^{LL})+\left(C_6-\frac{1}{2}C_8\right)(M_{\phi}^{SP}+M_{\pi\pi}^{SP})\Big\},
\\
\mathcal{A}_{B_s}(q\bar{q})&=&\frac{G_F}{2}\left\{V_{ub}^*V_{us}C_2M_{\pi\pi}^{LL}-V_{tb}^*V_{ts}
\left[\left(2C_4+\frac{1}{2}C_{10}\right)M_{\pi\pi}^{LL}+\left(2C_6+\frac{1}{2}C_8\right)M_{\pi\pi}^{SP}\right]\right\},
\\
\mathcal{A}_{B_s}(s\bar{s})&=&-\frac{G_F}{\sqrt{2}}V_{tb}^*V_{ts}\left\{\left(a_3+a_4+a_5-\frac{1}{2}(a_7-a_9-a_{10})\right)F_{\phi}^{LL}\right.
\nonumber\\
&&+\left(C_3+C_4-\frac{1}{2}(C_9+C_{10})\right)(M_{\phi}^{LL}+M_{\pi\pi}^{LL})+\left(C_5-\frac{1}{2}C_7\right)(M_{\phi}^{LL}+M_{\pi\pi}^{LR})
\nonumber\\
&&+\left(C_6-\frac{1}{2}C_8\right)(M_{\phi}^{SP}+M_{\pi\pi}^{SP})
  +\left(a_3+a_4+a_5-\frac{1}{2}(a_7+a_9+a_{10})\right)(A_{\phi}^{LL}+A_{\pi\pi}^{LL})\nonumber\\
&&+\left(a_6-\frac{1}{2}a_8\right)(A_{\phi}^{SP}+A_{\pi\pi}^{SP})+\left(C_3+C_4-\frac{1}{2}(C_9+C_{10})\right)(M_{\phi}^{LL}+M_{\pi\pi}^{LL})
\nonumber\\
&&\left.+\left(C_5-\frac{1}{2}C_7\right)(M_{\phi}^{LR}+M_{\pi\pi}^{LR})+\left(C_6-\frac{1}{2}C_8\right)(M_{\phi}^{SP}+M_{\pi\pi}^{SP})\right\}.
\end{eqnarray}
These $a_i(i=3,4,5,6,7,8,9,10)$ are the combined Wilson coefficients defined as
\begin{eqnarray}
a_i&=&C_i+\frac{1}{3}C_{i+1}\; \; i=3,5,7,9\\
a_i&=&C_i+\frac{1}{3}C_{i-1} \;\; i=4,6,8,10,
\end{eqnarray}
with $C_i$ are the Wilson coefficients. In this same way, we can also calculate the amplitudes of $B_{d,s}\to \phi (f_2\to)\pi^+\pi^-$, however we do not present them here due to the limited space. Last, we can obtain the differential branching fraction
\begin{eqnarray}
\frac{d^2\mathcal{B}}{d \zeta d\omega}=\frac{\tau\omega|\vec{p}_1||\vec{p}_3|}{32\pi^3 m_{B}^3}|\mathcal{A}|^2.
\end{eqnarray}
The magnitudes of three-momenta of one pion and the bachelor particle $\phi$ in the rest frame of the $\pi\pi$-pair are given by
\begin{eqnarray}
|\vec{p}_1|=\frac{\sqrt{\lambda(\omega^2,m_\pi^2,m_{\pi}^2)}}{2\omega}, \quad
|\vec{p}_3|=\frac{\sqrt{\lambda(m_{B}^2,m_\phi^2,\omega^2)}}{2\omega},
\end{eqnarray}
with the standard K$\ddot{a}$ll$\acute{e}$n function $\lambda (a,b,c)= a^2+b^2+c^2-2(ab+ac+bc)$.
\section{Numerical Results and Discussions}\label{sec:result}
We start this section with listing the parameters used in the numerical calculations, such as the mass of the mesons, the decay constants and the lifetimes of the $B$ mesons, the width of the intermediate resonances, the CKM matrix elements, and the QCD scale\cite{Zyla:2020zbs},
\begin{eqnarray}
&&m_{B}/m_{B_s}=5.279/5.366 ~{\rm GeV},\; m_{f_0}=(0.99\pm0.02)~{\rm GeV},\; m_{f_2}=1.275~{\rm GeV},\nonumber\\
&&V_{tb}=1.0,\;V_{td}=0.00854^{+0.00023}_{-0.00016},\;V_{ub}=0.00361^{+0.00011}_{-0.00009},\nonumber\\
&&V_{us}=0.22650\pm0.00048,\;V_{ts}=0.03978^{+0.00082}_{-0.00060},\;\Lambda_{QCD}^{f=4}=(0.25\pm0.05)~{\rm GeV},\nonumber\\
&&f_B=0.19\pm0.02~{\rm GeV},\;\;\;f_{B_s}=0.23\pm0.02~{\rm GeV}\nonumber\\
&& \Gamma_{f_2}=186 ~{\rm MeV},\,\, \tau_{B_d}/\tau_{B_s}=1.519/1.515~{\rm ps}.
\end{eqnarray}

Based on the obtained decay amplitudes in previous section and the input parameters above, we can calculate the $CP$-averaged branching fractions, the $CP$ asymmetries parameters, and the polarization fractions of final states for these considered $B_{d,s}\to \phi(f_{0,2}\to) \pi^+\pi^-$ decays. In Table.~\ref{br}, we present our results of the branching fractions, together with the currently available experimental measurements from the LHCb Collaboration \cite{Aaij:2016qnm}. We acknowledge that there are many uncertainties in our calculations. In this work we mainly take three kinds of errors into accounts, as shown in tables. The first errors are caused by parameters in the distribution amplitudes of the $B$ mesons, $\phi$ meson and the $\pi\pi$ pair, such as the shape parameter of $B_{(s)}$ meson $\omega/\omega_s=0.4\pm0.04/0.5\pm0.05$ GeV, the Gegenbauer moments of the $\phi$ meson distribution amplitudes, and the Gagenbauer moments $a_S$, $a_D^{(T)}$ corresponding to the $S$-wave and $D$-wave two-pion distribution amplitudes, respectively. The second kinds of errors arise from the higher order and higher power corrections, which are represented by varying the $\Lambda_{QCD}=(0.25\pm0.05)GeV$ and the factorization scale $t$ from $0.8t$ to $1.2t$. The last ones are from the uncertainties of the CKM matrix elements. From the table, we can see that the major uncertainties are the first ones, so we hope the future developments of the nonperturbative approaches such as the QCD sum rules and the Lattice QCD approach, can reduce these uncertainties.

From the Table.~\ref{br}, it is found that for the decays $B_s\to \phi \pi^+\pi^-$, our results could accommodate the current experimental results with large uncertainties, although for the $B_s \to \phi(f_0\to)\pi^+ \pi^-$ the theoretical cental value is twice of the experimental data. We also note that the branching fractions of $B_s\to \phi(f_{0,2}\to)\pi^+\pi^-$ decays are much larger than those of $B_d\to \phi(f_{0,2}\to)\pi^+\pi^-$ decays by three orders, it is mainly because the $B_d$ decays are suppressed by the CKM matrix elements $|{V_{td}}/{V_{ts}}|^2$. For $B_d\to \phi \pi^+\pi^-$, the branching fractions are at the order of ${\cal O}(10^{-9})$, which is hoped to be tested in the ongoing LHCb and Belle II experiments.

\begin{table}[!htb]
\caption{$CP$ averaged branching ratios (in $10^{-6}$) of $B_{d,s}\to \phi  (f_0/f_2 \to) \pi^+\pi^-$ decays in PQCD approach together
with experimental data from refs.\cite{Aaij:2016qnm,Zyla:2020zbs}. }
 \label{br}
\begin{center}
\begin{tabular}{l c c }
 \hline \hline
 \multicolumn{1}{c}{Decay Modes}&\multicolumn{1}{c}{PQCD } &\multicolumn{1}{c}{EXP}  \\
\hline\hline
 $B_s \to \phi(f_0\to)\pi^+ \pi^-$    &$2.35^{+3.17+2.50+0.44}_{-0.98-0.33-0.45}$  &$1.12\pm0.16^{+0.09}_{-0.08}\pm0.11$ \\

 $B_s \to \phi(f_2\to)\pi^+ \pi^-$   &$0.75^{+0.38+0.24+0.02}_{-0.32-0.21-0.03}$   &$0.61\pm0.13^{+0.12}_{-0.05}\pm0.06$ \\

 $B^0 \to \phi(f_0\to)\pi^+ \pi^-$  \;  &$2.97^{+1.92+1.63+2.74}_{-1.68-1.77-0.00}\times 10^{-3}$   &$<0.38 $  \\

 $B^0 \to \phi(f_2\to)\pi^+ \pi^- $ \;   &$2.46^{+1.73+0.37+0.29}_{-0.87-0.38-0.02}\times 10^{-3}$   &... \\

 \hline \hline
\end{tabular}
\end{center}
\end{table}

Unlike the quasi-two-body decays with the subprocess $f_0\to KK$ where the narrow-width approximation invalids, the narrow-width approximation is reliable in the decays with the subprocess $f_0\to \pi\pi$, which can be expressed as
\begin{eqnarray}
\mathcal{B}(B\to M(R\to)P_1P_2)\simeq\mathcal{B}(B\to MR)\times \mathcal{B}(R\to P_1P_2).
\end{eqnarray}
Under the above approximation and isospin relations, we can estimate the corresponding quasi-two-body decays $B\to \phi(f_0\to) \pi^0\pi^0$ within the experimental branching fractions of $B\to \phi(f_0\to)\pi^+\pi^-$. To achieve this goal, we first define the ratio $\mathcal{R}_1$ as
\begin{eqnarray}
\mathcal{R}_1&=&\frac{\mathcal{B}(B\to \phi(f_0\to)\pi^+\pi^-)}{\mathcal{B}(B\to \phi(f_0\to)\pi^0\pi^0)}\simeq\frac{\mathcal{B}(B\to\phi f_0)\times\mathcal{B}(f_0\to \pi^+\pi^-)}{\mathcal{B}(B\to\phi f_0)\times\mathcal{B}(f_0\to \pi^0\pi^0)}\nonumber\\
&\simeq&\frac{\mathcal{B}(f_0\to \pi^+\pi^-)}{\mathcal{B}(f_0\to \pi^0\pi^0)}=2,
\end{eqnarray}
Thereby, the branching fractions of $B_{d,s}\to \phi(f_0\to)\pi^0\pi^0$ can be obtained as
\begin{eqnarray}
\mathcal{B}(B_{d}\to \phi(f_0\to)\pi^0\pi^0)&=&\frac{\mathcal{B}(B_{d}\to \phi(f_0\to)\pi^+\pi^-)}{2}=(1.49_{-1.22}^{+1.86})\times 10^{-9},\\
\mathcal{B}(B_{s}\to \phi(f_0\to)\pi^0\pi^0)&=&\frac{\mathcal{B}(B_{s}\to \phi(f_0\to)\pi^+\pi^-)}{2}=(1.17_{-0.56}^{+2.03})\times10^{-6}.
\end{eqnarray}
Similarly, the branching fractions of the quasi-two-body $B_{d,s}\to \phi(f_2\to)\pi^+\pi^-$ decays are predicted to be
\begin{eqnarray}
\mathcal{B}(B_{s}\to \phi(f_2\to)\pi^0\pi^0)&=&\frac{\mathcal{B}(B_{s}\to \phi(f_2\to)\pi^+\pi^-)}{2}=(3.75_{-1.92}^{+2.24})\times10^{-7},\\
\mathcal{B}(B_{d}\to \phi(f_2\to)\pi^0\pi^0)&=&\frac{\mathcal{B}(B_{d}\to \phi(f_2\to)\pi^+\pi^-)}{2}=(1.23_{-0.47}^{+0.86})\times 10^{-9}.
\end{eqnarray}

Since the branching fraction of $f_0\to \pi\pi $ decay is still unknown, so we cannot extract the branching fractions of the corresponding two-body $B_{d,s}\to\phi f_0$ decays under the narrow-width approximation with the measured quasi-two-body decays. However, we could theoretically evaluate the ratio $\mathcal{R}_{2}$ between the $B_{s}\to\phi f_0$ and the $B_{d}\to\phi f_0$ decays using the narrow-width approximation
\begin{eqnarray}
\mathcal{R}_2&=&\frac{\mathcal{B}(B_{s}\to\phi f_0)}{\mathcal{B}(B_{d}\to\phi f_0)}\simeq\frac{\mathcal{B}(B_{s}\to\phi f_0)\times\mathcal{B}(f_0\to\pi^+\pi^-)}{\mathcal{B}(B_{d}\to\phi f_0)\times\mathcal{B}(f_0\to\pi^+\pi^-)}\nonumber\\
&\simeq&\frac{\mathcal{B}(B_{s}\to \phi(f_0\to)\pi^+\pi^-)}{\mathcal{B}(B_{d}\to \phi(f_0\to)\pi^+\pi^-)}\simeq8\times 10^2.
\end{eqnarray}
Once one of the $B_{s}\to\phi f_0$ and $B_{d}\to\phi f_0$ decays were measured in the future experiments, we could estimate the branching fraction of another decay. For the decays $B\to \phi f_2$, the branching fraction of $f_2\to \pi\pi$ decay has been measured to be $(84.2^{+2.9}_{-0.9})\%$\cite{Zyla:2020zbs}. Therefore, under the narrow width approximation, the branching fractions of $B_{d,s}\to \phi f_2$ decays are estimated to be
\begin{eqnarray}
\mathcal{B}(B_d \to \phi f_2)&\simeq&\frac{\mathcal{B}(B_d\to \phi (f_2\to)\pi^+\pi^-)}{\frac{2}{3}\times\mathcal{B}(f_2\to\pi\pi)}\simeq(4.38_{-1.69}^{+3.18})\times10^{-9},
\end{eqnarray}
\begin{eqnarray}
\mathcal{B}(B_s \to \phi f_2)&\simeq&\frac{\mathcal{B}(B_s\to \phi (f_2\to)\pi^+\pi^-)}{\frac{2}{3}\times\mathcal{B}(f_2\to\pi\pi)}\simeq(1.33_{-0.66}^{+0.80})\times10^{-6}.
\end{eqnarray}
In 2013, one of authors (Zou) has studied the branching fractions of the decays $B_{d}\to \phi f_2$ and $B_s \to \phi f_2$ \cite{Kim:2013cpa} in the PQCD approach, where $f_2$ was regarded as pure $q\bar{q}$ state. Compared current results and ones in ref.\cite{Kim:2013cpa}, these branching fractions are in agreement with each other with large uncertainties. The acceptable differences origin from the mixing angle between $q\bar q$ and $s\bar s$.

The branching fraction of $f_2\to K\overline{K}$ decay has been measured to be $(4.6^{+0.5}_{-0.4})\%$ \cite{Zyla:2020zbs}. Using the narrow width approximation, we also can evaluate the branching fractions of corresponding quasi-two-body $B \to \phi (f_2\to)K^+K^-$ decays. At first, we define the ratio $\mathcal{R}_{3}$ and calculate it as
\begin{eqnarray}
\mathcal{R}_3&=&\frac{\mathcal{B}(B\to \phi(f_2\to)\pi^+\pi^-)}{\mathcal{B}(B\to \phi(f_2\to)K^+K^-)}\simeq\frac{\mathcal{B}(B\to\phi f_2)\times \mathcal{B}(f_2\to\pi^+\pi^-)}{\mathcal{B}(B\to\phi f_2)\times \mathcal{B}(f_2\to K^+K^-)}\nonumber\\
&\simeq&\frac{\mathcal{B}(f_2\to\pi^+\pi^-)}{\mathcal{B}(f_2\to K^+K^-)}\simeq24.4.
\end{eqnarray}
So the branching fractions of $B_{d,s}\to \phi (f_2\to)K^+K^-$ decays can be obtained as
\begin{eqnarray}
\mathcal{B}(B_d\to \phi(f_2\to)K^+K^-)&=&\mathcal{B}(B_d\to \phi(f_2\to)\pi^+\pi^-)\frac{1}{\mathcal{R}_3}\nonumber\\
&\simeq&1.0\times 10^{-10},\\
\mathcal{B}(B_s\to \phi(f_2\to)K^+K^-)&=&\mathcal{B}(B_s\to \phi(f_2\to)\pi^+\pi^-)\frac{1}{\mathcal{R}_3}\nonumber\\
&\simeq&(3.07_{-1.56}^{+1.84})\times 10^{-8}.
\end{eqnarray}
We note that the branching fractions of $B_{d,s}\to \phi (f_0\to)K^+K^-$ cannot be obtained through the corresponding $B_{d,s}\to \phi (f_0\to)\pi^+\pi^-$ decays, because the narrow-width approximation is invalided in describing the line-shape of $f_0$, as discussed in previous section.

\begin{table}[!htb]
\caption{The direct $CP$ asymmetries parameters and the fractions of longitudinal polarization of final states (in $\%$) of $B_{d,s}\to \phi (f_0/f_2 \to) \pi^+\pi^-$ decays in PQCD approach. }
 \label{cpfl}
\begin{center}
\begin{tabular}{l c c }
 \hline \hline
 \multicolumn{1}{c}{Decay Modes}&\multicolumn{1}{c}{$A_{CP}^{dir}$ } &\multicolumn{1}{c}{$F_{L}$}  \\
\hline\hline
 $B_s \to \phi(f_0\to)\pi^+ \pi^-$    &$8.37_{-21.7-22.2-16.3}^{+22.2+14.7+17.6}$  &$...$ \\

 $B_s \to \phi(f_2\to)\pi^+ \pi^-$   &$4.45^{+2.66+0.29+1.70}_{-1.24-0.97-1.56}$   &$98.9_{-6.80-8.79-2.11}^{+1.43+2.8+0.00}$ \\

 $B^0 \to \phi(f_0\to)\pi^+ \pi^-$  \;  &0.0   &.... \\

 $B^0 \to \phi(f_2\to)\pi^+ \pi^- $ \;   &0.0  &$96.4_{-14.6-17.7-17.6}^{+1.83+0.00+0.00}$\\

 \hline \hline
\end{tabular}
\end{center}
\end{table}

At last, we turn to discuss the calculated $CP$ asymmetries and the fractions of longitudinal polarizations, which are given in Table.~\ref{cpfl}. It is known to us  the direct $CP$ asymmetry is proportional to the interference between contributions from the tree and penguin operators. Because the transitions $b\to s s \bar s$ and  $b\to s d \bar d$ are pure penguin processes, and $b\to s u \bar u$ is penguin dominated, so the $CP$ asymmetries of these decays are very small or even zero. Specifically, for the $B_d\to\phi(f_0/f_2\to)\pi^+\pi^-$ decays that are pure penguin processes, the direct asymmetries are zero. For the $B_s \to \phi(f_0/f_2\to)\pi^+\pi^-$ decays, the tree contributions are suppressed heavily by the CKM matrix elements, so the interference between the tree contribution and the penguin one is quite small, leading to small direct $CP$ asymmetries. From the Table.~\ref{cpfl}, we find that the $B_{d,s}\to\phi(f_2\to)\pi^+\pi^-$ decays are dominated by the longitudinal polarization contributions, which obeys the naive factorization assumption. The reason is that the contributions from transverse polarizations are power suppressed compared with these of longitudinal ones.
\section{Summary} \label{summary}
In this paper we investigated the resonant contributions of $B_{d,s}\to \phi\pi^+\pi^-$ decays with the $f_0(980)$ and $f_2(1270)$ as the intermediate resonant states, motivated by the recent measurements of LHCb collaboration. Within the $S$ wave and $D$ wave two-pion wave functions, we calculated the branching fractions, the $CP$ asymmetries, and the fractions of the longitudinal polarization. The obtained theoretical branching fractions of $B_s\to \phi(f_0(980)/f_2(1270)\to)\pi^+\pi^-$ decays are in agreement with the LHCb measurements within the errors. Furthermore, we also studied the $B_d \to\phi(f_0(980)/f_2(1270)\to)\pi^+\pi^-$, and the results can be further tested in the ongoing LHCb and Belle II experiments. The $CP$ asymmetries are small and even zero, because these concerned decays are either dominant by the penguin contribution or even pure penguin processes. For the $D$ wave decay channels, the fraction of the longitudinal polarization are very close to unity, which are in agreement with the naive factorization assumption. Based on the narrow width approximation, we also estimated the branching fractions of the $B_{d,s}\to \phi(f_2(1270)\to)\pi^0\pi^0$ and $B_{d,s}\to \phi(f_2(1270)\to)K^+K^-$ decays and the branching fractions of two-body decays $B_{d,s}\to \phi f_2(1270)$, which can be also be tested in the experiments.
\section*{Acknowledgment}
This work is supported in part by the National Science Foundation of China under the Grant Nos. 11705159, 11975195, 11875033, and 11765012, and the Natural Science Foundation of Shandong province under the Grant No. ZR2018JL001 and No.ZR2019JQ04. X. Liu is also supported by the Qing Lan Project of Jiangsu Province under Grant No. 9212218405, and by the Research Fund of Jiangsu Normal University under Grant No. HB2016004. This work is also supported by the Project of Shandong Province Higher Educational Science and Technology Program under Grants No. 2019KJJ007. Zhi-Tian Zou also acknowledges the Institute of Physics Academia Sinica for their hospitalities during the part of the work to be done.

\bibliographystyle{bibstyle}
\bibliography{mybibfile}
\end{document}